\documentclass[preprint2]{aastex}

\usepackage{amsmath}
\usepackage{graphicx}
\usepackage{epsfig,psfrag,epic,eepic}
\usepackage{textcomp}
\usepackage{times}
\usepackage{longtable}
\usepackage{lscape}

\slugcomment{To be submitted}

\shorttitle{Orbital motion of M-dwarfs}
\shortauthors{Janson et al.}

\begin{document}

\title{Orbital Monitoring of the AstraLux Large M-dwarf Multiplicity Sample\altaffilmark{*}}

\author{Markus Janson\altaffilmark{1,2,3}, 
Carolina Bergfors\altaffilmark{2,4}, 
Wolfgang Brandner\altaffilmark{2}, 
Micka{\"e}l Bonnefoy\altaffilmark{5}, 
Joshua Schlieder\altaffilmark{2}, 
Rainer K{\"o}hler\altaffilmark{2}, 
Felix Hormuth\altaffilmark{2}, 
Thomas Henning\altaffilmark{2}, 
Stefan Hippler\altaffilmark{2}
}

\altaffiltext{*}{Based on observations collected at the European Southern Observatory, Chile, under observing programs 081.C-0314(A), 082.C-0053(A), and 084.C-0812(A), and on observations collected at the Centro Astron\'omico Hispano Alem\'an (CAHA) at Calar Alto, operated jointly by the Max-Planck Institute for Astronomy and the Instituto de Astrof\'isica de Andaluc\'ia (CSIC).}
\altaffiltext{1}{Dept. of Astronomy, Stockholm University, Stockholm, Sweden; \texttt{markus.janson@astro.su.se}}
\altaffiltext{2}{Max Planck Institute for Astronomy, Heidelberg, Germany}
\altaffiltext{3}{Queen's University Belfast, Belfast, Northern Ireland, UK}
\altaffiltext{4}{University College London, London, UK}
\altaffiltext{5}{UJF-Grenoble, CNRS-INSU, IPAG, Grenoble, France}

\begin{abstract}\noindent
Orbital monitoring of M-type binaries is essential for constraining their fundamental properties. This is particularly useful in young systems, where the extended pre-main sequence evolution can allow for precise isochronal dating. Here, we present the continued astrometric monitoring of the more than 200 binaries of the AstraLux Large Multiplicity Survey, building both on our previous work, archival data, and new astrometric data spanning the range of 2010--2012. The sample is very young overall -- all included stars have known X-ray emission, and a significant fraction (18\%) of them have recently also been identified as members of young moving groups in the Solar neighborhood. We identify $\sim$30 targets that both have indications of being young and for which an orbit either has been closed or appears possible to close in a reasonable timeframe (a few years to a few decades). One of these cases, GJ~4326, is however identified as probably being substantially older than has been implied from its apparent moving group membership, based on astrometric and isochronal arguments. With further astrometric monitoring, these targets will provide a set of empirical isochrones, against which theoretical isochrones can be calibrated, and which can be used to evaluate the precise ages of nearby young moving groups.
\end{abstract}
%17.8% was the latest YMG count, but that does not include Schlieder etc., nor Malo 2012 (only Malo2014 and Kraus 2014)

\keywords{binaries: general --- techniques: high angular resolution --- stars: late-type}

\section{Introduction}
\label{s:intro}

Improvements in instrumentations and techniques for high-resolution imaging as well as other detection methods have benefitted the efficient study of large samples of stars with high sensitivity. This has led to a significant amount of multiplicity studies in recent years, spanning a wide range of stellar masses \citep[e.g.][]{burgasser2007,raghavan2010,kraus2011,janson2013,derosa2014}. One such technique that has become available over the past decade thanks to developments of visible light detectors with high-speed readout and low read noise is so-called Lucky Imaging \citep[e.g.][]{tubbs2002,law2006}. Lucky Imaging takes advantage of the fact that the wavefront distortion pattern of the atmosphere varies on timescales of order 10~ms, and that in some instances, the severity of this distortion is significantly smaller than in others. By acquiring a long series of very short exposures, and selecting only those frames in which the measured wavefront is relatively pristine, spatial resolution can be gained (at the cost of sensitivity), as a nearly diffraction-limited Point Spread Function (PSF) is reached. Previously, we have used this technique for an extensive high-resolution imaging survey of multiplicity in $\sim$700 early-to-mid M-dwarf stars \citep{bergfors2010,janson2012}, with a recent extension to late-type M-dwarfs \citep{janson2014}. These studies, along with other recent studies in the same spectral type range \citep[e.g.][]{law2008,dhital2010,dieterich2012,jodar2013} significantly expand the samples of previous surveys \citep[e.g.][]{fischer1992,reid1997,delfosse2004} and help to constrain the statistical distributions in M-dwarf multiplicity \citep[for a recent summary, see][]{duchene2013}.

However, these surveys do not only provide results on multiplicity statistics. The M-dwarfs multiples provided by the surveys can also be used as laboratories for constraining low-mass stellar properties. Since targets of the \citet{janson2012} study were chosen from the \citet{riaz2006} sample of X-ray-selected nearby M-dwarfs, many of them are quite young. Indeed, many ($\sim$150) of the targets have been identified as probable members of young moving groups in recent detailed studies \citep[e.g.][]{schlieder2012,malo2013,malo2014}. Young M-dwarf binaries for which dynamical masses can be determined are excellent for calibrating evolutionary models of M-dwarfs \citep[e.g.][]{burrows1997,baraffe1998}, particularly if the age is known. Conversely, to the extent that the evolutionary relationships are known, absolute ages can be determined isochronally in young M-dwarf binaries \citep[e.g.][]{janson2007}. Even regardless of theoretical models, M-dwarf binaries can give useful constraints on ages. For instance, a population of M-dwarf binaries with dynamically determined masses in young moving groups can be used to construct empirical isochrones, where relative ages between moving groups can be determined, and co-evality within groups can be stringently assessed. While extensive orbital monitoring campaigns have been performed among binaries in the M-type range \citep[e.g.][]{segransan2000,dupuy2010}, some of which are young; and while dedicated studies have been performed for some individual known young moving group members \citep[e.g.][]{bonnefoy2009,konopacky2007,kohler2013}, the sample of young low-mass binaries with well-characterized orbits remains small. Hence, it is of considerable importance to identify more such binaries for which an orbit can be closed over a reasonable time frame.

In this paper, we perform astrometric follow-up of binaries detected or confirmed in \citet{bergfors2010} and \citet{janson2012}, in order to confirm common proper motion in those cases where this has not yet been accomplished, and to better constrain their orbital properties. We also cross-check these binaries against identifications of members in young moving groups, in order to determine which systems are most promising for rigorously determining dynamical masses in the near future, and thus provide strong age constraints and/or model constraints for both the binaries themselves and the broader range of members in moving groups with which they may be associated. The paper is structured as follows: In Sect. \ref{s:obs}, we describe the observations that have been performed as part of this study, as well as the data reduction procedures. We then discuss the astrometric analysis in Sect. \ref{s:astro}, followed by the corresponding orbital analysis in Sect. \ref{s:orbits}, where we examine both the general sample as well as some particularly interesting individual cases. Finally, we summarize our results and conclusions in Sect. \ref{s:summary}.

\section{Observations and Data Reduction}
\label{s:obs}

This work is based primarily on observations acquired over several years with the AstraLux Norte camera \citep{hormuth2008} on the 2.2m telescope at Calar Alto in Spain, and the AstraLux Sur camera \citep{hippler2009} on the ESO/NTT 3.5m telescope at La Silla in Chile. The observations acquired specifically within the context of this program (orbital monitoring of M-dwarf binaries) were taken in Nov 2011, Jan 2012, and Aug 2012 with AstraLux Norte, and Oct 2010 (ESO program ID 086.C-0869(A)) and Jan 2012 (088.C-0753(A)) with AstraLux Sur. Additionally, some data taken for the purpose of our late-type M-dwarf sample \citep{janson2014} had a few targets overlapping with targets in this program, and so the astrometry from those data are included here as well. In some cases both $i^{\prime}$ and $z^{\prime}$ images were acquired and in some case only $z^{\prime}$ images, but here we will consistently only concern ourselves with the higher-quality $z^{\prime}$ data for the astrometric analysis. 

The observations used the same settings as are normally employed for AstraLux multiplicity observations \citep[e.g.][]{daemgen2009,bergfors2013}, with 15--30~ms individual readouts, adding up to 300~s of total integration time. For two of the Oct 2010 AstraLux Sur nights, the Barlow lens that was normally included in the light path was taken out. This led to a 2.8 times coarser pixel scale than normal, which has been taken into account in the astrometric calibration (see next section). Under normal circumstances, the full frame field of view of AstraLux Norte is a $\sim$24\arcsec\ square, and of AstraLux Sur is a $\sim$16\arcsec\ square, although subarray readouts are often employed to minimize readout times.

All in all, including both science and calibration observations, this study represents $\sim$500 new AstraLux observations of $\sim$10~min of telescope time each. In \citet{janson2012}, we presented 242 companion candidates, of which 219 were considered as probable real companions and 23 were probable background contaminants. Including the original observations, data published elsewhere in the literature, and the additional observations presented here, there are now two or more epochs of observation for 221 of these 242 candidates. Of these, 132 have three or more epochs of observations, 77 have four or more epochs, and 21 have five or more epochs. For a few of the targets, we have also recovered archival data that have not been previously published, these are discussed for those individual targets.

\begin{figure}[p]
\centering
\includegraphics[width=8cm]{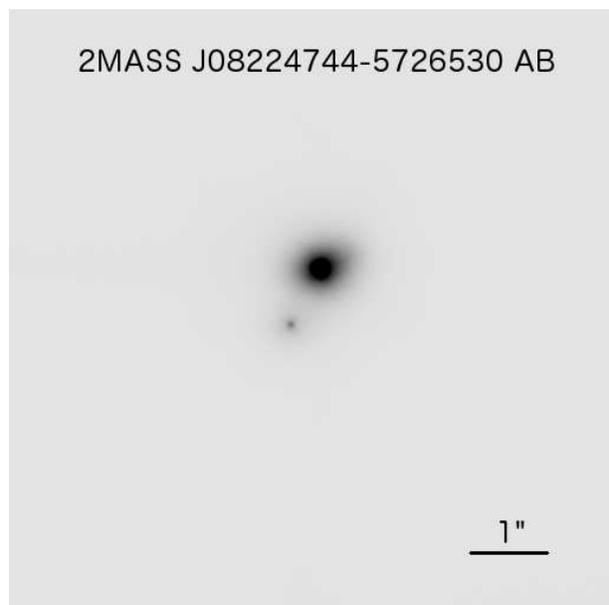}
\caption{Example image from the AstraLux campaign, showing the close AB pair of the 2MASS~J08224744-5726530 multiple system. A third component exists at a significantly wider separation (outside of the field of view of this observation). North is up and East is to the left in the image.}
\label{f:j0822im}
\end{figure}

Data reduction was performed with the pipeline described in \citet{hormuth2008}, to produce four collapsed frames per observation with the best 1\%, 2.5\%, 5\% and 10\% of the individual frames, respectively. In this study, we consistently use the 10\% selection for all targets. An example image from the campaign is shown in Fig. \ref{f:j0822im}.

% 77 = 76 from the table plus one for Seymour

\section{Astrometry}
\label{s:astro}

Astrometric measurements for the new data were acquired in different ways depending on the properties of the binary: for wide binaries, where the individual PSFs of the component stars were well separated, Gaussian centroiding was used to calculate relative astrometry in detector coordinates. For closer systems, the iterative PSF fitting scheme described in \citet{bergfors2010} was used. In many cases of binaries with components of nearly equal brightness, the well-known false triple effect that commonly occurs in Lucky Imaging data was evident \citep[see e.g.][]{law2006,bergfors2010}, such that three PSF components had to be fit simultaneously. All PSF fitting was done with three different reference star PSFs in order to evaluate the uncertainty of the procedure and the impact of PSF variability.

For astrometric calibration, we observed Trapezium whenever it was visible, and M15 otherwise. Reference astrometry was acquired from \citet{mccaughrean1994} in the case of Trapezium and \citet{marel2002} for M15. We measured Gaussian centroids for five of the brightest stars in each astrometric field and referenced their relative separations and position angles to the literature values, and also compared the results to the IRAF \textit{geomap} procedure used in previous work \citep[e.g.][]{kohler2008,bergfors2010}. It was found that the pixel scale could vary by up to 1\% depending on the choice of reference stars and methods, hence this is consistently used as the error in pixel scale. The uncertainty in field rotation was found to be 0.3$^{\rm o}$. In this way, we determined the pixel scale and angle of true North of AstraLux Sur as 15.18~mas/pixel and 1.49$^{\rm o}$ for Oct 2010 with the Barlow lens in the beam; 42.83~mas/pixel and 2.09$^{\rm o}$ for Oct 2010 without the Barlow lens; and 15.16~mas/pixel and 2.36$^{\rm o}$ for Jan 2012. In order to acquire a consistently calibrated astrometry over all epochs, we also re-calculate the astrometric calibration of the Nov 2008 epoch as 15.17~mas/pixel and 1.90$^{\rm o}$, and of the Jan 2010 epoch as 15.18~mas/pixel and 2.51$^{\rm o}$. For AstraLux Norte, the corresponding values are 23.56~mas/pixel and 1.66$^{\rm o}$ for Nov 2011; 23.58~mas/pixel and 1.72$^{\rm o}$ for Jan 2012; and 22.59~mas/pixel and 1.83$^{\rm o}$ for Aug 2012. Here, the angle of true North is defined as the counter-clockwise orientation of true North with respect to the y-axis -- in other words, in calculating a true position angle in sky coordinates, a positive angle of true North is subtracted from the position angle measured in detector coordinates. 

We also carried out an analysis of potential higher order geometric distortions of the AstraLux cameras by comparing AstraLux Norte observations of M13 with Hubble Space Telescope (HST) images. The HST images were processed by the standard HST pipeline, including corrections for geometric distortions. The analysis used 25 single and relatively isolated stars in common between AstraLux and HST observations, distributed over the AstraLux field of view. After correction for difference in the image scales and instrumental position angles, no systematic residuals could be identified. AstraLux image scales in the x- and y-coordinate directions are identical within the measurement uncertainties. For AstraLux Norte, the RMS values for the positional residuals of the stars are of the order of 10~mas to 15~mas over the full 24\arcsec\ by 24\arcsec\ field of view. These residuals amount to about 1/20 of the Full Width at Half Maximum (FWHM) of the PSF for data that were obtained in a night with 1.6 arcsec seeing, which is a typical PSF centroiding accuracy under such circumstances. Over these scales, the errors in pixel scale and orientation derived above amount to 60--120~mas, hence these latter factors dominate the error budget, rendering higher-order geometric distortions negligible.

Common proper motion (CPM) was tested for all targets observed in multiple epochs, by extracting proper motions from the NOMAD catalogue\footnote{Available from \url{http://www.nofs.navy.mil/nomad/} or VizieR I/297}, or when unavailable there, from the PPMXL catalogue \citep{roeser2010}. The expected background trajectory was calculated for each of these systems using these proper motions along with estimations of the system parallax. For this purpose, we used measured trigonometric parallaxes whenever available, and estimated parallaxes from the photometric distances otherwise. In order for a companion candidate to be considered as having a demonstrated CPM, it had to deviate from the background trajectory by at least 3$\sigma$ in at least one epoch. To further count as showing signficant signs of orbital motion, it had to also deviate from its original relative astrometry by at least 3$\sigma$. This method is robust for confirming CPM and thus validating physical companion candidates, but it is not strictly robust for rejecting CPM, since orbital motion can in principle mimic non-CPM (i.e., the orbital motion brings the companion close to the background trajectory by chance). We discussed some limit cases in \citet{janson2012}, and continually mark such cases with `U' for `Unclear' in the CPM assessment, along with cases for which the errors are simply too large with respect to the system proper motion for any significant evaluation to be performed. In many cases, we have been able to confirm CPM and orbital motion that were not yet confirmed in \citet{janson2012}, though in several cases these properties remain undetermined due to slow proper motion or orbital motion. In a few cases (e.g. J13015919+4241160 and J14430789+1720463), it is even the case that a CPM that was previously regarded as significant is now deemed unclear; this is however simply due to the fact that we adopt more conservative astrometric calibration errors in this study. Our CPM results confirm the notion that the vast majority of binary candidates in \citet{janson2012} are indeed physical companions. The astrometric values from this study as well as from the literature for the full set of multi-epoch targets are listed in Table \ref{t:multep}.

% Refer to relevant tables/figures

\section{Orbital constraints}
\label{s:orbits}

For many of the multiple systems in the survey, no meaningful orbital constraints can be imposed at this point. These are targets that either have only been observed in two epochs, or for which the orbital motion is too small over the observational baseline to cover any reasonable fraction of the orbit. However, there are many targets for which it appears feasible to close an orbit in a foreseeable future, and some for which an orbital closure has already been accomplished.  Some cases of particular interest are discussed in the following sub-sections. In Table \ref{t:shortperiod}, we provide qualitative assessments of the orbital properties of all systems that have an estimated period $P_{\rm est}$ of less than 50 years. $P_{\rm est}$ was estimated in \citet{janson2012} purely on the basis of the estimated mass of the system and with the assumption that its projected separation is close to its semi-major axis. In addition to $P_{\rm est}$, we include a general assessment of the orbital speed, where 'rapid' implies that it seems plausible that an orbit could be closed in less than $\sim$40~years if continued at the current rate, 'intermediate' implies slower motion but still indicative of a $<$100~yr orbit, and 'slow' implies a slower orbit still. We keep these as purely indicative and qualitative assessments for now, due to the large uncertainties remaining in many cases. For instance, a binary displaying a slow motion over a few year baseline could be highly eccentric and presently residing near apastron, and may thus turn out to close an orbit on much shorter timescales than an extrapolation of the current motion would imply. For the cases where an actual orbital fit could be made, we used the procedure developed \citet{kohler2008,kohler2012}. 

Aside from the orbital assessments in Table \ref{t:shortperiod}, we also make brief individual comments, including if the star has been identified as a young moving group (YMG) member. For this purpose, we use primarily \citet{malo2013} and \citet{malo2014} to identify YMG candidates but we also run the BANYAN II online tool \citep{gagne2014} in order to test the impact of priors on the estimated membership probabilities. \citet{malo2013} use uniform prior probabilities for membership in the various kinematic groups and the field, while BANYAN II allows for the use of more conservative priors that take into account the fact that only a small fraction of local stars should be members of young associations. The tool also allows for adjusting the priors based on whether there are separate indications of a young ($<$1~Gyr) age for a given object. All cases marked `YMG' with no question mark in the table have $>80$\% YMG membership probabilities regardless of the choice in priors. Cases with question marks are more complex and are discussed individually in subsequent sections. 

As mentioned in \citet{janson2012}, J16552880-0820103 and J22232904+3227334 have already had detailed orbital studies in the past, and we refer to the corresponding articles \citep{segransan2000,seymour2002} for discussion of these individual cases.

\subsection{2MASS J00325313-0434068}
\label{s:j0032}

\begin{figure*}[p]
\centering
\includegraphics[width=16cm]{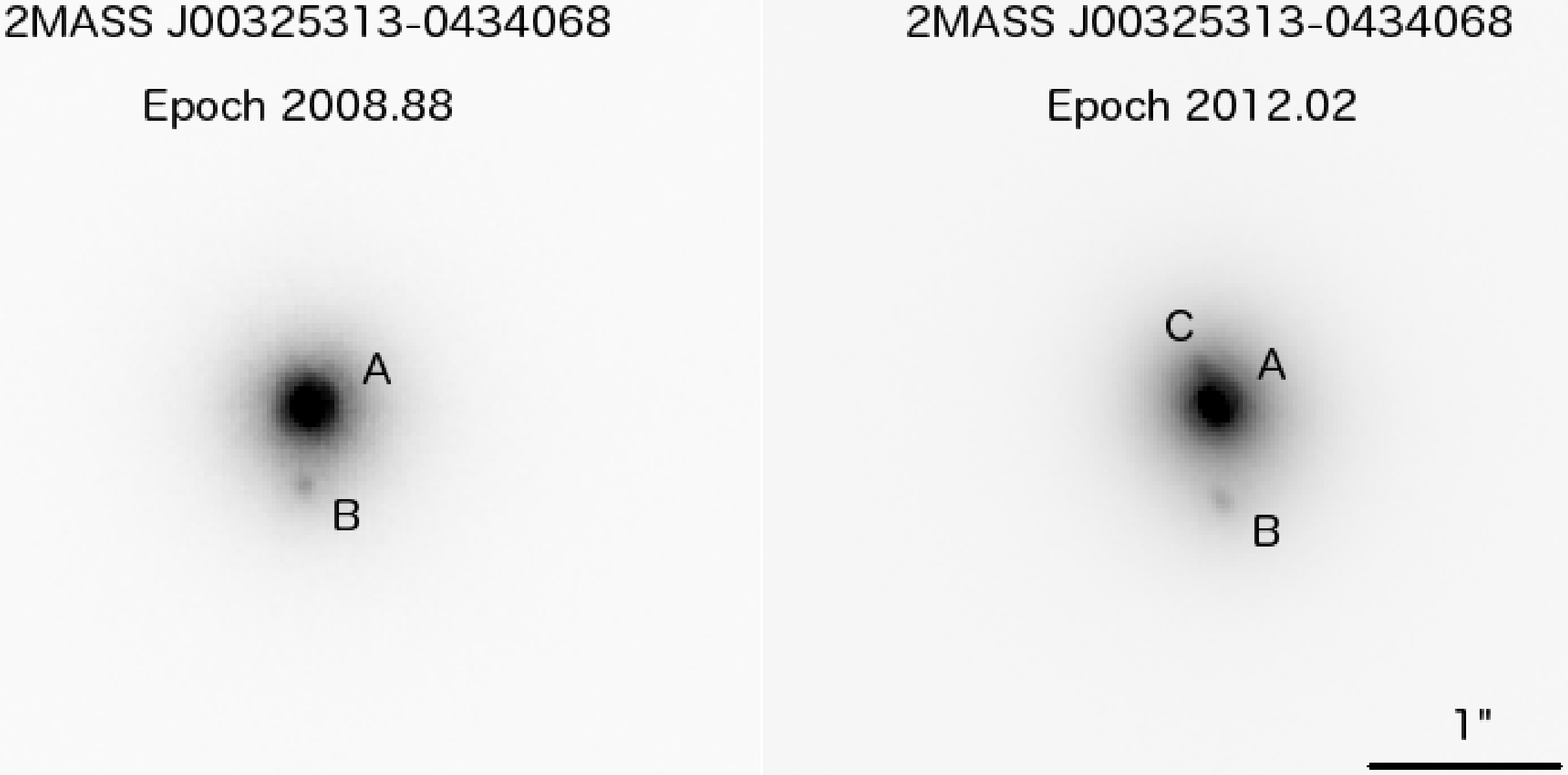}
\caption{Two epochs of imaging for the 2MASS J00325313-0434068 triple system. The `C' component is invisible in late 2008 (left), but emerges above the brighter `A' component in early 2012 (right). North is up and East is to the left in the images.}
\label{f:j0032combo}
\end{figure*}

The AB pair of 2MASS J00325313-0434068 appears to be moving rather slowly, but in our third epoch image, an additional component C is seen in the images (see Fig. \ref{f:j0032combo}). The fact that it is located at $\sim$200~mas from the primary component in 2012 but was invisible in 2008 could imply a quite rapid orbital motion outwards in the meantime, so this could be an interesting system for determining orbital parameters for at least the AC pair in a short timescale.

\subsection{2MASS J01112542+1526214}
\label{s:j0111}
% Lots of data points, MG

\citet{malo2013} identified this target as a probable member of the $\beta$~Pic moving group, and it is included as a bona fide member of this group in \citet{malo2014}. As originally discovered by \citet{beuzit2004}, the star is a binary, and it has been observed in eight additional epochs with AstraLux Sur and Norte. Hence, there are a lot of observational data spanning 12 years and covering substantial orbital motion. However, the orbit is long enough that no tight constraints can be placed on it yet. Orbital periods of 30--40 years give excellent fits to the data, hence it is certainly plausible that a substantial fraction of the orbit can be covered in the next decade, but considerably longer orbits also give reasonable fits, so it remains too early to say whether signficiant dynamical constraints on the evolutionary stage of this binary can be deduced in that timeframe. It is, in any case, a highly relevant target for further monitoring in the future.

\subsection{2MASS J01433150+3904165}
\label{s:j0143}
% Implied short orbit

The apparent orbital motion of this binary is very rapid, implying perhaps an orbital period of order one decade. However, only three data points exist so far, and in addition, the binary consists of two components of similar brightness, and undergoes the false triplet effect, such that it becomes difficult, particularly in the third epoch, to determine which component is which. Thus, there is a 180$^{\rm o}$ phase ambiguity for that data point, and more data will be needed to resolve it. It is therefore premature to place any constraints on the orbit, but it is clearly a very interesting case for the future, with additional observations.

\subsection{2MASS J02132062+3648506}
\label{s:j0213}
% Covering a complete orbit (but not much in between)

\begin{figure}[p]
\centering
\includegraphics[width=8cm]{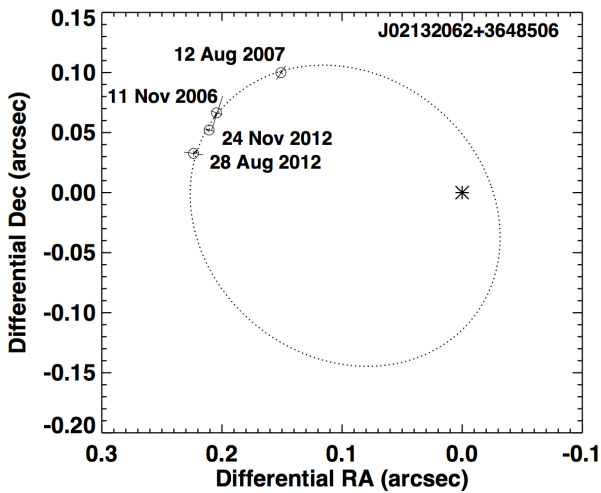}
\caption{An example of an orbital fit to the binary 2MASS J02132062+3648506. By chance, the available data points cover one section of the orbital phase, over one full orbit, which means that the orbital period can be well constrained, but very little can be determined about any of the other orbital parameters.}
\label{f:j0213orbit}
\end{figure}

By chance, the period of this binary orbit is almost exactly equal to the dominant time span of the observational baseline. Two pairs of data points have been acquired where the span within each pair is less than one year, and the span of the full baseline is about 6 years, which is also close to the period of the binary. As a result, the data nearly closes a full orbit but doesn't cover much in between, so while good constraints can be placed on the orbital period, essentially no constraints can be drawn otherwise. The 67\% confidence limits for orbital period on well-fitting orbits are 6.13--7.15 years, corresponding to a precision of $\sim$8\%. This is a very promising target for acquiring a highly constraining fit with just a few addditional data points. A plot with an example orbital fit is shown in Fig. \ref{f:j0213orbit}.

\subsection{2MASS J02165488-2322133}
\label{s:j0216}
% BC pair is new, too close in previous epochs

The AB pair of this multiple system is wide, as was reported in \citet{bergfors2010}. However, our follow-up of the system has revealed a much closer companion to the B component, which due to the small separation is a probable physical companion, but the astrometry is still insufficient to test this. The companion is moving outwards and was simply too close to the B component in the 2008 epoch to be resolved. Even in the 2010 and 2012 epochs, the detection is rather marginal and consequently the astrometric errors are large, hence more time will be required before common proper motion and orbital motion can be assessed.

\subsection{2MASS J02271603-2929263}
\label{s:j0227}
% Removed bad data point

One data point from \citet{janson2012} for J02271603-2929263 gives a clearly deviant astrometric value; visual re-inspection shows that this is due to a very poor quality of the image, hence we have discarded that data point for this study.

\subsection{2MASS J03323578+2843554}
\label{s:j0332}
% Very peculiar orbit, MG

2MASS J03323578+2843554 was first identified as a probable $\beta$~Pic member in \citet{schlieder2012}, and is given a 99.9\% membership probability in \citet{malo2014}. Our analysis using BANYAN II confirms a high probability of membership, although it can be as low as 82.6\% if we choose non-uniform priors and do not take into account the fact that there are separate indications of a $<1$~Gyr age for the system. We reported that the target is a triple system in \citet{janson2012}. With our new epoch of imaging, the BC pair shows a clear continuous orbital motion with respect to component A. The motion within the BC pair itself is a more complicated matter. In \citet{janson2012}, we noted that the scatter in separation was larger than the error bars would imply, in a manner that seemed random and incompatible with orbital motion. The early astrometric data points for the system were derived with a separate procedure from the other data points, hence we have re-reduced the full set of data with a homogenous procedure and a uniform calibration scheme. With the new reduction, the data is much better consistent with orbital motion within the error bars. The motion is such that there is very little observable motion during 2006--2009, but then substantial motion occurs for the 2012 data point, where the projected separation is smaller and the position angle approximately the same. This implies a high-eccentricity orbit, close to an edge-on orientation. Since a rather limited portion of the orbit appears to be covered so far, we cannot impose strict constraints on the orbital elements, but it seems plausible that the orbital period might be short, with periods down to $\sim$10~yr providing good fits to the data.

\begin{figure}[p]
\centering
\includegraphics[width=8cm]{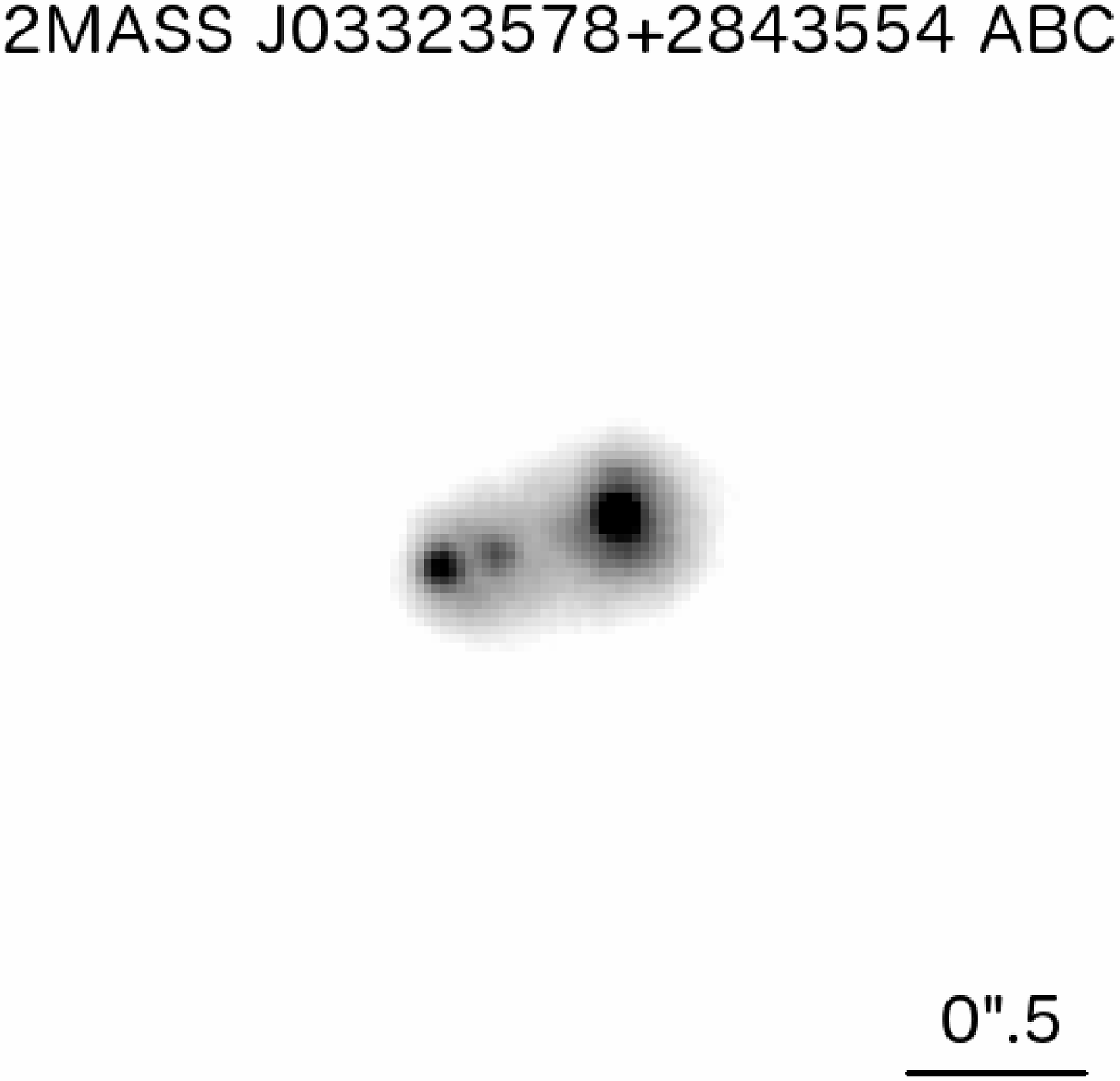}
\caption{Image of the 2MASS J03323578+2843554 triple system from Feb 2009, with the A component to the right, the B component to the left and the C component in the middle. North is up and East is to the left in the image. }
\label{f:j0332im}
\end{figure}

Interestingly, the three components line up neatly with each other (see Fig. \ref{f:j0332im}); for instance, in the 2012 epoch, the differential angle of two lines along the axes of the AB pair and the BC pair respectively is only 3.2$^{\rm o}$. The direction of the BC motion vector from 2006 to 2012 is essentially indistinguishable from the 2012 axis of the pair, and the direction of the AB motion vector over the same timeframe differs from the 2012 AB axis by only $\sim$11$^{\rm o}$. This possibly implies a high degree of coplanarity for the motion of the inner and outer pair of the 2MASS J03323578+2843554 system, where both pairs are close to edge-on. Meanwhile, a collective analysis of high-order multiple systems in the field shows no preference toward co-planarity between any set of pairs in the system \citep{janson2010}. With further study of 2MASS J03323578+2843554 and other young high-order multiples, it would become possible to test for whether the relative orientations in such systems are age dependent, following dynamical evolution of the system.

\subsection{2MASS J04373746-0229282 (GJ 3305)}
\label{s:j0437}
% Implied short orbit, MG

GJ 3305 has a very large number of astrometric data points both with AstraLux and in the literature, and is furthermore identified as a bona fide member of the $\beta$~Pic moving group. A dominant fraction of the orbit is now starting to be covered. The new astrometry presented in this work is summarized in Table \ref{t:multep}. A detailed orbital analysis of this orbit, including also radial velocity data, will be presented in Bonnefoy et al. (in prep.), along with near-infrared spectra of both components.

\subsection{2MASS J04595855-0333123}
\label{s:j0459}
The quite rapidly moving binary 2MASS J04595855-0333123 was identified as a possible member of the Argus moving group in \citet{malo2013}. Our analysis with BANYAN II confirms an extremely high membership probability (99.9\%) in the case of uniform priors, but the probability drops to 27.7\% if non-uniform priors are chosen. However, using non-uniform priors combined with a $<1$~Gyr age gives a probability of 68.8\% for Argus membership. Since there are indeed indications from X-ray emission that this sample is young \citep{janson2012}, we consider it more likely than not that 2MASS J04595855-0333123 is an Argus member, but we note that the contamination probability is substantial.

\subsection{2MASS J05241914-1601153}
\label{s:j0524}
While the estimated period of 2MASS J05241914-1601153 from its projected separation is $\sim$40~yr, its slow orbital motion between late 2008 and early 2012 implies a significantly longer period, unless the current orbital configuration is close to apastron. There is noticable motion only in the radial direction, suggesting a close to edge-on orbit. Both \citet{malo2013} and our independent BANYAN II check give a high $\beta$~Pic moving group membership probability ($>$99\%) for this system if using uniform priors. However, with non-uniform priors this probability drops down to 26.7\%. Even if we consider the fact that it is probable that the age of the star is $<1$~Gyr, the probability is still only 29.6\%. Hence, the dominant probability states that 2MASS J05241914-1601153 is kinematically unrelated to $\beta$~Pic and other known moving groups, though a finite probability remains of it being an actual member.

\subsection{2MASS J05320450-0305291}
\label{s:j0532}
Between 2009 and 2012, the binary 2MASS J05320450-0305291 moves along at a modest rate relative to what the separation-based period estimate of $\sim$23~yr would imply, though with significant motion in both the radial and azimuthal directions. Although our BANYAN II analysis confirms the proposition in \citet{malo2013} that 2MASS J05320450-0305291 is a very high-probability member of the $\beta$~Pic moving group if uniform priors are chosen, the probability is substantially smaller if more realistic, non-uniform priors are selected. Nominally, the membership probability is 12.4\% for non-uniform priors, increasing only marginally to 15.8\% if a $<1$~Gyr age is assumed. It thus seems significantly more likely that this system is unrelated to the moving group, although actual membership remains a faint possibility that cannot be very stringently excluded at this point.

\subsection{2MASS J06134539-2352077}
\label{s:j0613}
% Implied short orbit, MG

Estimated as a high-probability ($>$99\%) member of the Argus moving group in \citet{malo2014}, which is confirmed with BANYAN II analysis, this binary is an excellent candidate for providing age constraints and/or tests of evolutionary models over a reasonable time frame. The binarity was discovered in \citet{janson2012}, although only a single epoch was available at the time. Our follow-up in this work confirms common proper motion, and displays rapid orbital motion. In only just under two years, the binary has moved by more than 40$^{\rm o}$ in position angle. With only two epochs available over a short baseline, it is of course impossible to constrain the orbital period as of yet, but our best estimate implies $\sim$25 years.

\subsection{2MASS J06161032-1320422}
\label{s:j0616}

A slow outward motion from late 2008 to early 2012 indicates that the orbital period of 2MASS J06161032-1320422 may be longer than estimated from its separation ($\sim$37 years). \citet{malo2013} identified this target as a probable member of the $\beta$~Pic moving group. Our BANYAN II check confirms a 99.6\% probability of membership if uniform priors are assumed -- however, this drastically drops when adopting non-uniform priors. In this circumstance, the probability of membership is a mere 3.6\%, increasing only subtly to 4.4\% if adopting a $<1$~Gyr age. This implies that the kinematic similarity of 2MASS J06161032-1320422 to the $\beta$~Pic moving group is coincidental.

\subsection{2MASS J07174710-2558554}
\label{s:j0717}
% Implied short orbit, not the one seen in Bergfors

The binary pair that is discussed here is in fact not the one that was originally discussed for this system in \citet{bergfors2010}. In that previous publication, a faint object near the detection limit was noted which is a probable background star, but which has not been recovered in subsequent epochs. However, in the 2010 and 2012 epochs presented here, another component is seen, which is a close companion that shares a common proper motion with the primary. The binary is very close and moving outward, so it was not noticed in 2008 where the separation is only 80~mas, but going back to the 2008 data, we have been able to recover the companion also in that epoch. The binary motion is fairly rapid, implying that the orbit can be closed over a few decades.

\subsection{2MASS J07285137-3014490 (GJ 2060)}
\label{s:j0728}
% GJ 2060

Since it has both been identified as a bona fide member of the AB Dor moving group \citep[e.g.][]{malo2013} and has a rich amount of orbital motion spanning several epochs and closing the orbit, GJ 2060 is a prime target for examining mass-luminosity relationships in young binaries. One issue that is important to note for this target is that while the epoch is quoted as 2002.83 for the literature value in \citet{daemgen2007}, the actual epoch is 2005.83, as we have verified by locating the archival data in the Gemini Science Archive. We have also retreived and analyzed archival data from Keck/NIRC2 taken in very late 2002. We use the header astrometry for nominal astrometric values, and adopt errors as given for NIRC2 in \citet{konopacky2007}. The values are listed in Table \ref{t:multep}. A detailed analysis of the orbit and near-infrared spectra for the individual components of this system will be presented in Bonnefoy et al. (in prep.).

\subsection{2MASS J09053033-4918382}
\label{s:j0905}
% The wrong star was probably observed in epoch 1

Our 2010.82 and 2012.01 epochs of this system show a drastically different system architecture from the 2010.08 image. The 2010.08 image presented in \citet{janson2012} shows a binary candidate of highly unequal brightness components, at a separation of $\sim$4\arcsec. On the other hand, the 2010.82 and 2012.01 epochs both show a nearly equal brightness binary with a separation of $\sim$270~mas, with statistically significant common proper motion and orbital motion. There is no sign of any faint and far candidate in the latter epochs, and no sign of any bright and close companion in the former epoch. The solution most probably lies in the fact that 2MASS J09053033-4918382 by chance is located only 17\arcsec\ away from the unrelated F-type star 2MASS J09052993-4918544, which has almost the same near-infrared brightness. It appears probable that the stars were mixed up in the first epoch of observation, and that the F-type star was observed instead of the intended M-type star. Indeed, looking back into the header astrometry, the coordinates are closer to the F-type star than the M-type star in the first epoch. We thus tentatively assume that the 2010.82 and 2012.01 epochs correctly image 2MASS J09053033-4918382, and that the 2010.08 epoch does not, which changes the parameters of the binary from \citet{janson2012}. However, it would be desirable in a future observation to systematically observe 2MASS J09053033-4918382 and 2MASS J09052993-4918544 in sequence, to verify that this interpretation is indeed the correct one.

\subsection{2MASS J10364483+1521394}
\label{s:j1036}
% Implied short orbit for the BC pair

As for the case of 2MASS J03323578+2843554, the early astrometric data points  in \citet{janson2012} for this triple system were based on an obsolete procedure, hence we have re-analyzed all of the astrometry with a uniform procedure. It is clear that rapid orbital motion is occuring for the BC pair. However, since the B and C components have almost identical brightnesses, there are 180$^{\rm o}$ ambiguities in the astrometry, and it is unclear if the Jan 2008 data can be trusted since those data are of particularly poor quality. We thus consider that more astrometric points will be necessary before a detailed orbital fit is meaningful.

%\subsection{2MASS J16552880-0820103}
%\label{s:j1655}
% Segransan orbit

%As we already noted in \citet{janson2012}, this is one case for which a full orbit is covered (in both imaging and radial velocity), and orbital parameters have already been calculated to good precision in \citet{segransan2000}.

\subsection{2MASS J20163382-0711456}
\label{s:j2016}

The orbital motion of 2MASS J20163382-0711456 is quite rapid, with approximately 30$^{\rm o}$ azimuthal motion and 60\% outward radial motion in three years. The system was identified as a probable Argus association member in \citet{malo2013}. Our independent check with BANYAN II confirms this: if assuming uniform priors, the probability is 99.8\%. This drops to 77.6\% when adopting non-uniform priors, which is still quite high probability. Furthermore, constraining the age to $<1$~Gyr while maintaining non-uniform priors bumps the probability back up to 84.7\%. Thus, we conclude that the system is a probable member of Argus, though at a lower level of confidence than the clearest cases.

\subsection{2MASS J23172807+1936469 (GJ~4326)}
\label{s:j2317}
% Covering a full orbit, good constraints, probably not MG after all

\begin{figure}[p]
\centering
\includegraphics[width=6cm]{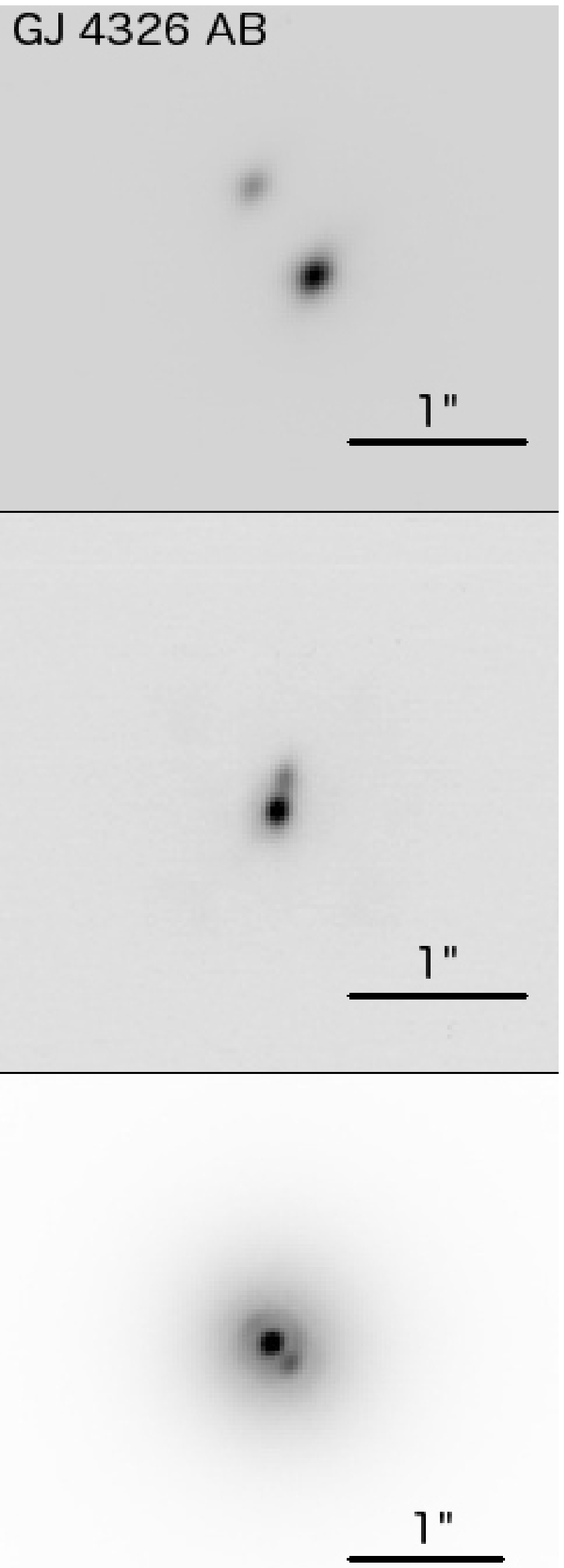}
\caption{Three of the images that exist for the GJ~4326 system. North is up and East is to the left in each case. The top and middle images are archival NACO images from 2004 and 2010, respectively. The bottom image is the newest AstraLux image from 2012.}
\label{f:j2317series}
\end{figure}

\begin{figure}[p]
\centering
\includegraphics[width=8cm]{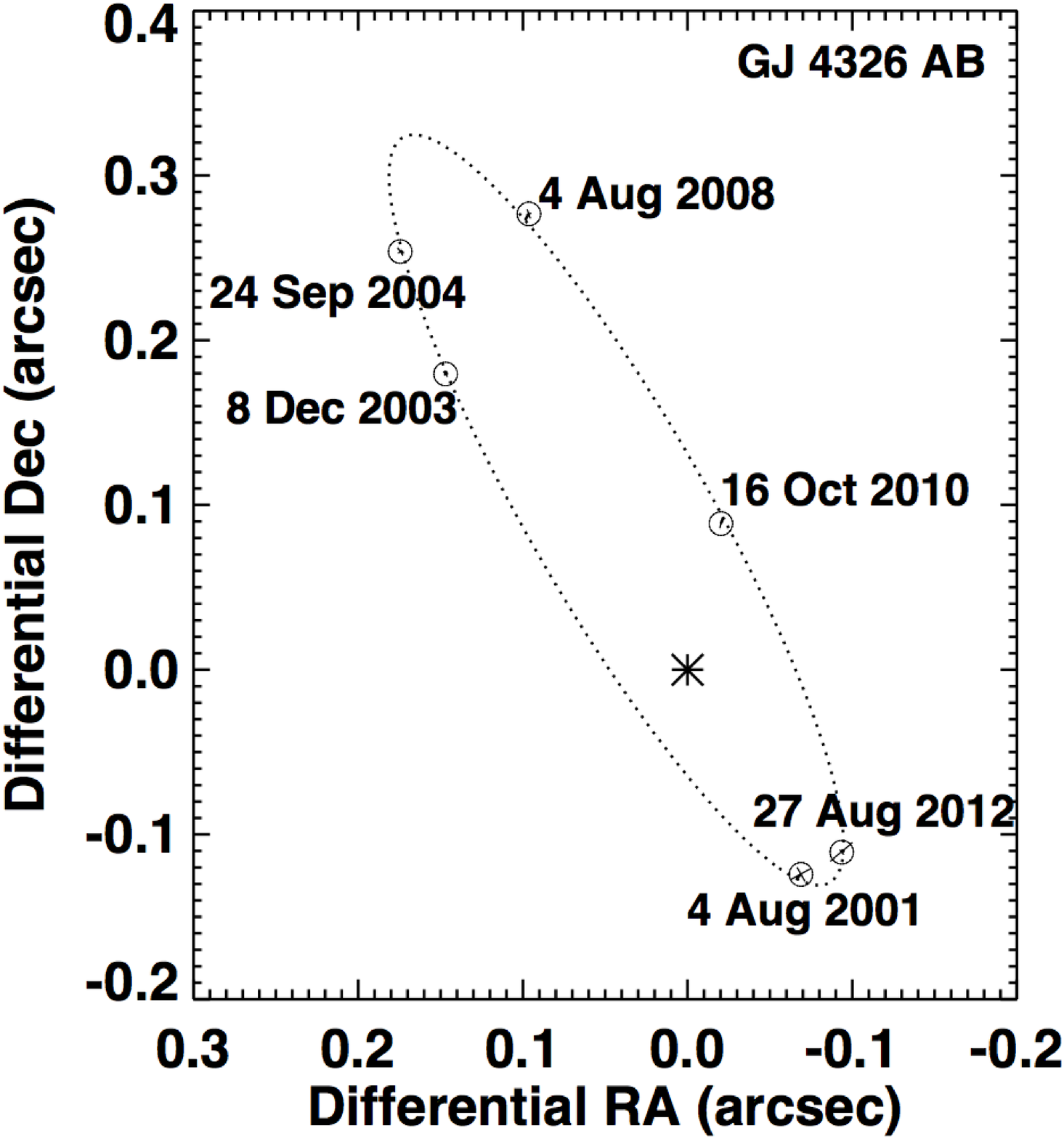}
\caption{Orbital fit to the full set of resolved images that exist for the GJ 4326 system. The data covers a wide range of orbital phases and almost close an orbit, such that good constraints can already be imposed on all orbital parameters (apart from the linear semimajor axis, since the distance to the system is associated with a large uncertainty.)}
\label{f:j2317orbit}
\end{figure}

GJ~4326 is one of the cases in the sample in which a rather high-quality orbit can be determined already with the given set of data. Given that it has been identified as a member of the $\beta$~Pic moving group with a 94.4\% probability in \citet{malo2013}, this makes it a prime interest target for mass-luminosity analysis (see below for further discussion on membership probabilities). The binary was first resolved by \citet{beuzit2004} in 2000, and we have an additional two epochs from AstraLux in 2008 and 2012 which by themselves would not provide a strong estimate on the orbit. However, three archival epochs of VLT/NACO \citep{lenzen2003,rousset2003} observations are also present from different programs (072.C-0570, 073.C-0155, and 086.C-0515)  in 2003, 2004, and 2010. We have analyzed those data, adopting the header astrometry and assuming calibration errors of 0.3$^{\rm o}$ in true North orientation and 1\% in pixel scale. Taken together, the six data points both close an orbit, and cover it nicely along its various phases (see Fig. \ref{f:j2317series} and Fig. \ref{f:j2317orbit}). We have fit the orbit as described above, and the resulting orbital parameters are listed in Table \ref{t:astro4326}. If the distance to the target was perfectly known, a dynamical mass could be determined to a 2\% precision based on these parameters. Unfortunately, the distance estimate is 11.6$\pm$2.4~pc \citet{lepine2005}, which is a 21\% error. This has an enormous impact on the mass due to the $m \sim a^3$ scaling, so the actual lower and upper limits (67\% interval) are 0.1~$M_{\rm sun}$ and 0.3~$M_{\rm sun}$ respectively, which corresponds to a 50\% error. It is clear that a better distance will be required to stringently constrain this system. This will be provided by GAIA \citep{perryman2001}. 

\begin{table*}[p]
\caption{Orbital parameters of GJ 4326 AB}
\label{t:astro4326}
\centering
\begin{tabular}{ll}
\hline
\hline
Parameter & Value \\
\hline
$P$	&	$11.56^{+0.08}_{-0.10}$~yr	\\
$\alpha$	&	264$\pm$1~mas	\\
$e$	&	$0.464^{+0.004}_{-0.003}$	\\
$i\tablenotemark{a}$	&	103.3$\pm$0.2~deg	\\
$t_{\rm peri}$	&	$56419^{+3}_{-2}$~d	\\
$\omega$	&	-153.2$\pm$0.5~deg	\\
$\Omega\tablenotemark{a}$	&	30.8$\pm$0.2~deg	\\
\hline
\end{tabular}
\tablenotetext{a}{There is an ambiguity between $i$ and $180-i$ due to the lack of radial velocity information.}
\end{table*}

However, based on the data that we have, we can already draw some broad conclusions regarding the age of the system. There is a broad range of high-quality (unresolved) photometry for this system: It is listed in the SDSS \citep{ahn2012} which provides precise $u^{\prime}g^{\prime}r^{\prime}i^{\prime}z^{\prime}$ photometry, and there is 2MASS $JHK$ \citep{skrutskie2006} and WISE $W1$--$W4$ \citep{wright2010} photometry with small errors as well. Furthermore, there exist resolved images in the $i^{\prime}z^{\prime}H$-bands from AstraLux and NACO. These photometric values are summarized in Table \ref{t:phot4326}. We use a grid of BT-SETTL models \citep{allard2014} for an assumed age of 20~Myr \citep[evolutionary constraints provided by][]{baraffe1998} in order to try to fit these values. This is done by selecting every possible pairing of a primary mass with a secondary mass. Masses between grid points are interpolated using shape-preserving piecewise cubic interpolation. For each component of a pair, the BT-SETTL models then provide predictions of fluxes in every relevant photometric band, and the sums and differences of those fluxes are compared to the actual measured values. We find that no pair coupling of any properties for the individual components in the models can reproduce the data. When the total flux across the full wavelength range is matched, then the models are too red to fit the data. Correspondingly, in order to fit the color of the observations, the models necessarily predict a far too high total brightness (by a factor of at least 4). Another way to phrase it is that the effective temperatures and bolometric luminosities of the components do not match up consistently for the theoretical radii expected for objects at this age.

\begin{table}[p]
\caption{Photometry of GJ 4326 AB}
\label{t:phot4326}
\centering
\begin{tabular}{ll}
\hline
\hline
Band & Unresolved photometry \\
\hline
$u^{\prime}$	&	15.798$\pm$0.008~mag	\\
$g^{\prime}$	&	13.028$\pm$0.001~mag	\\
$r^{\prime}$	&	11.492$\pm$0.001~mag	\\
$i^{\prime}$	&	10.175$\pm$0.001~mag	\\
$z^{\prime}$	&	9.735$\pm$0.001~mag	\\
$J$	&	8.020$\pm$0.024~mag	\\
$H$	&	7.411$\pm$0.020~mag	\\
$K$	&	7.173$\pm$0.017~mag	\\
$W1$	&	7.001$\pm$0.034~mag	\\
$W2$	&	6.862$\pm$0.021~mag	\\
$W3$	&	6.779$\pm$0.016~mag	\\
$W4$	&	6.663$\pm$0.060~mag	\\
\hline
Band & $\Delta$ mag \\
\hline
$i^{\prime}$	&	1.70$\pm$0.15~mag	\\
$z^{\prime}$	&	1.50$\pm$0.12~mag	\\
$H$	&	1.18$\pm$0.12~mag	\\
\hline
\end{tabular}
\end{table}

There are a few possible reasons for why this situation might come about. For instance, we might hypothesize that the system is much more distant than estimated. If the distance was two times the estimated 11.6~pc value, then a situation would emerge in which the true luminosities of the components are four times higher, providing the possiblity for the models to reproduce the data. However, apart from deviating from the existing distance estimate by 4.8~$\sigma$, which seems excessive, such a view is also inconsistent with the orbital data. If the distance were made 2 times larger, the semi-major axis would increase by the same factor, and the mass required to match the given period at this semi-major axis would increase by a factor 8. This would correspond to super-Solar masses, which can be firmly excluded from the unresolved spectral type of the system (as well as from its colors and the luminosity match to the models). The only remaining way to reach a fully self-consistent picture is to reduce the radii by a factor 2 relative to the nominal model predictions. Unless the models are drastically incorrect, this can only be accomplished by increasing the age of the system. Once the binary components have reached the main sequence, which happens at around 100~Myr, a self-consistent fit involving all factors dicussed here can be reached. After this point, M-dwarfs are essentially fixed on the main sequence, so any arbitrary $>$100~Myr age is allowed in this sense.

As we noted above, GJ~4326 has an estimated 94.4\% probability of being a member of the $\beta$~Pic moving group in \citet{malo2013}, with no update in \citet{malo2014}. Running the online BANYAN II tool \citep{gagne2014} gives probabilities of 91.4\% with default priors, and 98.6\% if we constrain the age to younger than 1~Gyr, both of which are comparable to the original estimate. Most of the remaining probability, 1.4--8.6\%, is for it being an unrelated field object. While this latter probability is small, it is not zero. More data, and in particular a better constrained distance, will greatly benefit the analysis of this system. However, until the advent of such data, we consider it likely that GJ~4326 is a rare kinematic imposter in the $\beta$~Pic moving group, and is an unrelated, older object. It could still be quite young (down to $\sim$100~Myr), but hardly young enough to match the estimated moving group age range of $\sim$10--20~Myr \citep[e.g.][]{zuckerman2001,binks2014}.

\subsection{2MASS J23495365+2427493}
\label{s:j2349}

With a 42$^{\rm o}$ azimuthal motion over just 3 years, 2MASS J23495365+2427493 might close its orbit a bit faster than its separation based period estimate of 38~years would indicate. The system was identified as an ambiguous moving group member candidate in \citet{malo2013}, in that its motion could conceivably fit both to the $\beta$~Pic and Columba moving groups. Our BANYAN II check broadly agrees with this: assuming uniform priors, there is a 94.8\% match probability to $\beta$~Pic and a 5.1\% match probability to Columba (with the remaining probability assigned to the field hypothesis). If non-uniform priors are assumed, the Columba match probability becomes negligible, and the $\beta$~Pic probability becomes 47.1\%. However, if the assumption that the age is $<1$~Gyr is adopted, the $\beta$~Pic value rises back up to 80.3\% and the Columba hypothesis becomes marginally feasible again at 2.7\%. It thus seems sensible to classify 2MASS J23495365+2427493 primarily as a probable $\beta$~Pic moving group member.

\section{Summary and Conclusions}
\label{s:summary}

In this paper, we have analyzed more than 500 new images acquired for astrometric follow-up of the AstraLux M-dwarf multiplicity survey. With these data, we have been able to confirm common proper motion for several binaries for which this was previously not possible, confirming that indeed the vast majority of candidates in the survey, and particularly among those that pass photometric criteria, are real physical companions. Furthermore, the orbital information built up for some of the binaries is now substantial, allowing for tight constraints on their orbital parameters already now or in the near future. This is particularly relevant for those targets that have been identified as being members of nearby young moving groups. In this paper, we have identified more than 10 binaries that are both plausible (or probable) YMG members and have estimated orbital periods of less than $\sim$40~yr -- in some cases just a few years. These will be excellent targets for isochronal analysis of moving group ages, particularly once GAIA has provided precise distances to all of them, which is currently a severely limiting quantity for many analytic purposes. One target that has been analyzed in particular detail here is GJ~4326, which has been assigned a 94.4\% probability of being a member of the $\beta$~Pic moving group in the literature. Our orbital and photometric analysis however suggests that the binary components are inconsistent with an age as young as a membership of this moving group would imply, with radii more reminiscent of M-dwarfs on the main sequence, and therefore indicating that the binary is a low-probability kinematic imposter to the group. While a more precise distance estimate will be required before anything more definitive can be said about the system, this illustrates the utility of isochronal analysis of suspected YMG members. Likewise, our studies of other young binaries that have not (as of yet, at least) been identified as members of any kinematic association can lead to a better understanding of their age, which is otherwise difficult to determine accurately. In addition to all these issues, continued orbital monitoring of this binary sample will be important for acquiring better estimates for the distribution of orbital parameters, such as the actual semi-major axis distribution (rather than one based on projected separations) and the eccentricity distribution.

\acknowledgements
We thank all the staff at the Calar Alto and La Silla observatories for their support. M.J. gratefully acknowledges support from the Knut and Alice Wallenberg Foundation. This study made use of the CDS services SIMBAD and VizieR, as well as the SAO/NASA ADS service and the archival services for the VLT, Gemini, Keck, and Subaru telescopes. This publication makes use of data products from the Wide-field Infrared Survey Explorer, which is a joint project of the University of California, Los Angeles, and the Jet Propulsion Laboratory/California Institute of Technology, funded by the National Aeronautics and Space Administration.

\appendix

{\scriptsize
\begin{longtable}{lrrrrrrlll}
\caption{Relative astrometry of all stellar pairs of this study.}\\
\hline\hline
2MASS ID & Sep. & PA & Epoch & $d$ & $\mu_{\rm RA}$ & $\mu_{Dec}$ & Ref$^{\rm a}$ & CPM & OM \\
 & (\arcsec) & (deg) & (yr) & (pc) & (mas/yr) & (mas/yr) & & & \\
\hline
\endfirsthead
\caption{continued.}\\
\hline\hline
2MASS ID & Sep. & PA & Epoch & $d$ & $\mu_{\rm RA}$ & $\mu_{Dec}$ & Ref$^{\rm a}$ & CPM & OM \\
 & (\arcsec) & (deg) & (yr) & (pc) & (mas/yr) & (mas/yr) & & & \\
\hline
\endhead
\hline
\endfoot
\hline
\multicolumn{10}{l}{\scriptsize{$^a$Reference to original publication of the astrometric epoch. The values in the table will sometimes differ slightly from the values in the reference, }}\\
\multicolumn{10}{l}{\scriptsize{due to an updated astrometric calibration.}}\\
\hline
\endlastfoot
J00063925-0705354	&	0.230$\pm$0.006	&	6.5$\pm$0.5	&	2008.63	&	12	&	-108	&	94	&	\citet{janson2012}	&	Y	&	Y	\\
J00063925-0705354	&	0.247$\pm$0.003	&	5.3$\pm$0.4	&	2008.88	&	12	&	-108	&	94	&	\citet{janson2012}	&	Y	&	Y	\\
J00063925-0705354	&	0.336$\pm$0.004	&	1.9$\pm$0.3	&	2012.01	&	12	&	-108	&	94	&	This paper	&	Y	&	Y	\\
J00063925-0705354	&	0.322$\pm$0.004	&	1.9$\pm$0.3	&	2012.65	&	12	&	-108	&	94	&	\citet{janson2014}	&	Y	&	Y	\\
J00063925-0705354	&	0.337$\pm$0.003	&	1.4$\pm$0.3	&	2012.90	&	12	&	-108	&	94	&	\citet{janson2014}	&	Y	&	Y	\\
J00150240-7250326	&	0.286$\pm$0.009	&	68.9$\pm$0.3	&	2008.86	&	41	&	-116.5	&	-136.9	&	\citet{janson2012}	&	Y	&	Y	\\
J00150240-7250326	&	0.286$\pm$0.003	&	51.6$\pm$0.3	&	2010.82	&	41	&	-116.5	&	-136.9	&	This paper	&	Y	&	Y	\\
J00250428-3646176	&	0.597$\pm$0.012	&	242.2$\pm$0.3	&	2008.86	&	29	&	264.1	&	84	&	\citet{janson2012}	&	Y	&	Y	\\
\end{longtable}
}

{\scriptsize
\begin{longtable}{lrrll}
\caption{Assessments of orbital motion in systems with $P_{\rm est} < 50$~yr.}\\
\hline\hline
Target ID & Epochs$^{\rm a}$ & $P_{\rm est}$ (yr)$^{\rm b}$ & Motion$^{\rm c}$ & Comments$^{\rm d}$ \\
\hline
\endfirsthead
\caption{continued.}\\
\hline\hline
Target ID & Epochs$^{\rm a}$ & $P_{\rm est}$ (yr)$^{\rm b}$ & Motion$^{\rm c}$ & Comments$^{\rm d}$ \\
\hline
\endhead
\hline
\endfoot
\hline
\multicolumn{5}{l}{\scriptsize{$^a$Number of epochs of observation for the target.}}\\
\multicolumn{5}{l}{\scriptsize{$^b$Estimated orbital period from the projected separation and assumed system mass \citep{janson2012}.}}\\
\multicolumn{5}{l}{\scriptsize{$^c$Qualitative assessment of the orbital period from astrometric motion. `Rapid' implies a high probability}}\\
\multicolumn{5}{l}{\scriptsize{for the orbit to close within 40~yr.}}\\
\multicolumn{5}{l}{\scriptsize{$^d$Other comments, noting cases that have only been observed in a single epoch, have a too short baseline}}\\
\multicolumn{5}{l}{\scriptsize{for a useful assessment of orbital speed to be made, or are members of young moving groups (YMGs).}}\\
\hline
\endlastfoot
J00063925-0705354	&	5	&	14	&	slow 	&	---	\\
J00325313-0434068AB	&	3	&	25	&	slow 	&	---	\\
J00325313-0434068AC	&	1	&	9	&	unknown	&	single epoch	\\
J01112542+1526214	&	9	&	12	&	rapid	&	YMG	\\ 
J01154885+4702259	&	2	&	48	&	intermediate?	&	short baseline	\\
J01212520+2926143	&	2	&	46	&	rapid?	&	short baseline	\\
J02132062+3648506	&	4	&	8	&	very rapid 	&	---	\\
J02133021-4654505	&	3	&	7	&	rapid	&	---	\\
\end{longtable}
}

\end{document}